# Non-canonical propagation of high-order elliptic vortex-beams in a uniaxial crystal

T. Fadeyeva, C. Alexeyev, B. Sokolenko, M. Kudryavtseva and A. Volyar

*Physics Department, Taurida National V.I. Vernadsky University Vernadsky av.4, Simferopol, Ukraine, 95007. e-mail: volyar@crimea.edu*

**Summary**. We have derived the corresponding equations and found their solutions both for nonparaxial and paraxial beams. The paraxial solutions we have presented in the form of the generalized Hermite-Gaussian beams propagating perpendicular to the optical axis of a uniaxial crystal. We have also constructed the generalized Laguerre-Gaussian beams at the *z=0* plane and analyzed their evolution in a homogeneous isotropic medium. Comparing it with the evolution of the standard Laguerre-Gaussian beams with $n=0$ and $m \neq 0$ in the crystal we have revealed that the additional elliptic deformation of the extraordinary beam results in topological reactions that essentially distorts field structure for the account of different rotation rates of the vortex row originated from the centered degenerate optical vortex and the conoscopic pattern. We have predicted conversion of the vortex topological charge at the beam axis similar to that in astigmatic lenses and analyzed the radical differences with this process. We have revealed the synchronic oscillations of the spin angular momentum and the sign of the vortex topological charge at the beam axis.

## I. Introduction

A traditional description of a broad paraxial beam propagating orthogonally to the crystal optical axis is as a rule restricted by frameworks of the plane wave approach. Each plane wave is splintered into ordinary and extraordinary ones [1] at the crystal input. Since the phase velocities of these waves are different the initial polarization state of the wave is periodically transformed. More exact analysis have shown that the extraordinary Gaussian beam in the crystal has an elliptic shape whose form changes gradually when transmitting the beam while the ordinary beam experiences only scale transformations [2-4].

A circularly polarized singular beam bearing the optical vortex [5] (or simply a vortex-beam) brings its own correction into the propagation process. For example, the vortex beam whose axis is slightly tilted to the direction orthogonal to the crystal optical axis rotates when rotating the crystal axis, the optical vortex rotating separately from the host beam taking part in a complex precessional and nutational motions. This unique vortex property permits to create the optical reducer for the devices of micro-particle trapping, rotation and transportation [6]. But the principle point is that the above processes do not break the structural stability of the singular beam i.e. the vortex composition of the beam is permanent when propagating. Such a fact has been shown up strange from our point of view. Indeed, on one hand, different field distributions in the ordinary and extraordinary beams must results inevitably to the spatial beam depolarization; on the other hand, the different scales along the *x*- and *y*-axes must causes also distortion of the wavefronts of the partial beams provoking the phase perturbations in vicinity of the vortex core in each circularly polarized component and inducing, ipso facto, transformations of the centered vortex composition.

Thus, the aim of our paper is to analyze the propagation of the singular beam with elliptical cross-section bearing high-order optical vortices estimating the structural transformations of the vortex core and the vortex-beam as a whole

## II. Basic equations and their solutions

*1. Non-paraxial approach*

Let us consider the wave motion of the monochromatic circularly polarized singular beam propagating perpendicular to the optical axis of the unbounded non-absorbing uniaxial crystal (Fig.1) with the permittivity tensor:

$$\hat{\varepsilon} = \begin{pmatrix} \varepsilon_1 & 0 & 0 \\ 0 & \varepsilon_2 & 0 \\ 0 & 0 & \varepsilon_1 \end{pmatrix}. \qquad (1)$$

The Maxwell equations

$$\nabla \times \mathbf{E} = -ik\,\mathbf{H}, \quad \nabla \times \mathbf{H} = ik\,\mathbf{D}, \quad \mathbf{D} = \hat{\varepsilon}\,\mathbf{E} \qquad (2)$$

permit one to find solutions for the waves in the anisotropic medium with the material tensor (1) in the form of two beam types: *E*-beams and *H*-beams.

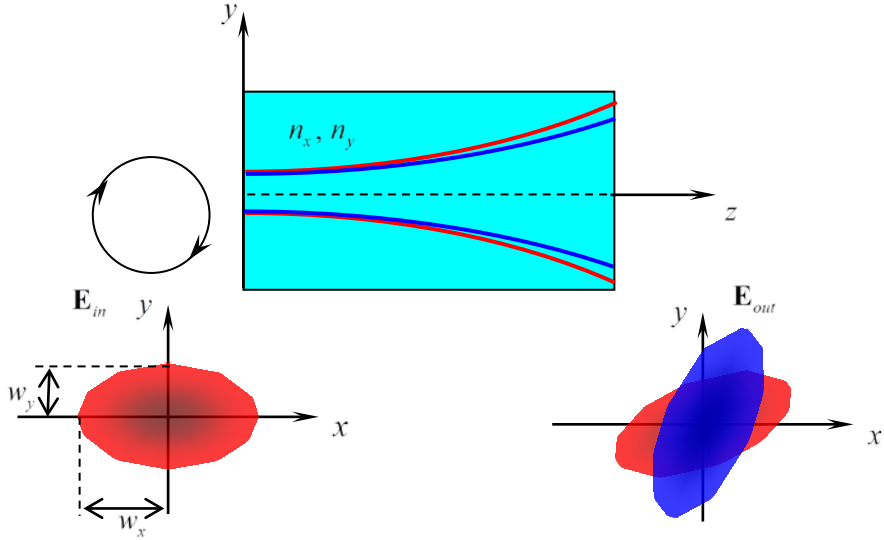

Fig. 1 Sketch of the elliptic beam propagation in a uniaxial crystal

*E-beams*

Form the equations (2) one can write the wave equation for the electric field $\mathbf{E}$ in the form

$$\nabla^2 \mathbf{E} + k^2\,\hat{\varepsilon}\,\mathbf{E} = \nabla(\nabla \mathbf{E}), \qquad (3)$$

where $k$ stands for the wave number in free space, and also the equation:

$$\varepsilon_1(\partial_x E_x + \partial_z E_z) + \varepsilon_2 \partial_y E_y = 0, \qquad (4)$$

the last equation being written as

$$\nabla \mathbf{E} = -\frac{\varepsilon_2 - \varepsilon_1}{\varepsilon_1} \partial_y E_y. \tag{5}$$

Supposing the $E_y$ component to vanish $E_y = 0$ we come to the equations:

$$\nabla^2 E_x + k^2 \varepsilon_1 E_x = 0, \quad \nabla^2 E_z + k^2 \varepsilon_1 E_z = 0 \tag{6}$$

and also

$$\partial_x E_x = -\partial_z E_z. \tag{7}$$

The equation (7) can be reduced to the identity if

$$E_x = \partial_z \Phi_E, \quad E_z = -\partial_x \Phi_E \tag{8}$$

while two equations (6) are reduced to the single scalar Helmholtz equation for the function $\Phi_E$:

$$\nabla^2 \Phi_E + k^2 \varepsilon_1 \Phi_E = 0.. \tag{9}$$

Finally, we can trace out the solutions of the Maxwell equations for E-beams (or ordinary beams) in the form

$$E_x = \partial_z \Phi_E, \tag{10} \quad H_x = \frac{i}{k} \partial^2_{xy} \Phi_E, \tag{13}$$

$$E_y = 0, \tag{11} \quad H_y = \frac{i}{k}\left(\partial^2_z + \partial^2_x\right)\Phi_E, \tag{14}$$

$$E_z = -\partial_x \Phi_E; \tag{12} \quad H_z = -\frac{i}{k}\partial^2_{yz}\Phi_E. \tag{15}$$

In fact, the solutions (10)-(15) represent the vector field of the non-paraxial beam propagating in a homogeneous isotropic medium with the refractive index $n_x = \sqrt{\varepsilon_1}$.

*H-beams*

Form the equations (2) we can also write the wave equation for the magnetic field $\mathbf{H}$ in the form

$$\nabla^2 \mathbf{H} + k^2 \varepsilon_1 \mathbf{H} = -(\varepsilon_1 - \varepsilon_2)(\mathbf{e_x}\partial_z - \mathbf{e_z}\partial_x)(\partial_z H_x - \partial_x H_z), \tag{16}$$

$$\partial_x H_x + \partial_z H_z + \partial_y H_y = 0. \tag{17}$$

Assuming the $H_y$ component to vanish $H_y = 0$ we obtain

$$\nabla^2 H_x + k^2 \varepsilon_1 H_x = -\frac{(\varepsilon_1 - \varepsilon_2)}{\varepsilon_2} \partial_z (\partial_z H_x - \partial_x H_z), \tag{18}$$

$$\nabla^2 H_z + k^2 \varepsilon_1 H_z = \frac{(\varepsilon_1 - \varepsilon_2)}{\varepsilon_2} \partial_x (\partial_z H_x - \partial_x H_z) \tag{19}$$

and also

$$\partial_x H_x = -\partial_z H_z. \tag{20}$$

Substituting eq (20) into eqs (18) and (19) we find that

$$\partial^2_x H_x + \frac{\varepsilon_2}{\varepsilon_1}\partial^2_y H_x + \partial^2_z H_x + k^2 \varepsilon_2 H_x = 0, \tag{21}$$

$$\partial_x^2 H_z + \frac{\varepsilon_2}{\varepsilon_1}\partial_y^2 H_z + \partial_z^2 H_z + k^2\varepsilon_2 H_z = 0. \tag{22}$$

These equations are consistent provided that

$$H_x = \partial_z \Phi_H, \quad H_z = -\partial_x \Phi_H \tag{23}$$

while the scalar function $\Phi_H$ obeys the equation:

$$\partial_x^2 \Phi_H + \frac{\varepsilon_2}{\varepsilon_1}\partial_y^2 \Phi_{Hx} + \partial_z^2 \Phi_H + k^2\varepsilon_2 \Phi_H = 0. \tag{24}$$

Thus, the electric and magnetic field of the *H*-beams (or extraordinary beams) are

$$E_x = -\frac{i}{k}\partial_{xy}^2 \Phi_H, \tag{25} \qquad H_x = \partial_z \Phi_H, \tag{28}$$

$$E_y = -\frac{i}{k}\left(\partial_z^2 + \partial_x^2\right)\Phi_H, \tag{26} \qquad H_y = 0, \tag{29}$$

$$E_z = \frac{i}{k}\partial_{yz}^2 \Phi_H; \tag{27} \qquad H_z = -\partial_x \Phi_H. \tag{30}$$

Thus, the non-paraxial *H*-beam propagates through the uniaxial crystal as if through a homogeneous isotropic medium with the refractive index $n_y = \sqrt{\varepsilon_2}$. However, such a medium has the particular property – it has different scales along the *x*- and *y*-axes, namely, $\Phi_H = \Phi_H(\bar{x}, \bar{y}, z)$, so that $\bar{x} = x$, $\bar{y} = \frac{n_x}{n_y} y$. This circumstance enables us to use the results obtained in Ref. [7] for nonparaxial ordinary and extraordinary beams that propagate along homogeneous isotropic media taking into account refractive indices $n_x$ and $n_y$ and also different space scaling for extraordinary beams. Nevertheless, laying aside the non-paraxial approach we will focus our attention on the paraxial vector beams.

## 2. Paraxial approximation for the E- and H-beams

Let the beam propagates along the *z*-direction in such a way that $\Phi_{E,H} = \Psi_{E,H}(x,y,z)\exp(ikn_{x,y}z)$ and the demand $\left|\partial_z^2\Psi\right| \ll k\left|\partial_z\Psi\right|$ is fulfilled. Then the equations (9) and (24) for the *E*- and *H*-beams is rewritten in the form

$$\partial_x^2 \Psi_E + \partial_y^2 \Psi_E + i2k_x\partial_z\Psi_E = 0, \quad k_x = kn_x \tag{31}$$

$$\partial_x^2 \Psi_H + \frac{n_y^2}{n_x^2}\partial_y^2 \Psi_H + i2k_y\partial_z\Psi_H = 0, \quad k_y = kn_y. \tag{32}$$

The electric and magnetic fields of the *E*- and *H*-paraxial beams get the form

*E-beam*

$$E_x = ik_x\Psi_E e^{ik_xz}, \tag{33} \qquad H_x \approx 0, \tag{36}$$

$$E_y = 0, \tag{34} \qquad H_y \approx -i\frac{k_x^2}{k}\Psi_E e^{ik_xz}, \tag{37}$$

$$E_z = -\frac{k_x}{k}\partial_x\Psi_E e^{ik_xz}; \tag{35} \qquad H_z \approx \frac{k_x}{k}\partial_y\Psi_E e^{ik_xz}. \tag{38}$$

*H-beams*

$$E_x \approx 0, \quad (39) \qquad H_x = ik_y \Psi_H e^{ik_y z}, \quad (42)$$

$$E_y \approx -i\frac{k_y^2}{k}\Psi_H e^{ik_y z}, \quad (40) \qquad H_y = 0, \quad (43)$$

$$E_z = -\frac{k_y}{k}\partial_y \Psi_H e^{ik_y z}; \quad (41) \qquad H_z = -\partial_x \Psi_H e^{ik_y z}. \quad (44)$$

### III. Elliptic beams

#### 1. Basic relationships

Since the initial circular cross-section of the *H*-beam at the plane $z=0$ turns into the elliptical one when transmitting through the crystal [3,4,6] we will find the solutions to the paraxial equations (31) and (32) in the form of elliptic beams. The theory of the elliptic and astigmatic beams in free space was developed and improved in a number of papers (see e.g. [8-14]). The general analysis of the generalized Hermite-Gaussian and Hermite-Laguerre-Gaussian beams in free space were considered in the papers [15-18]. Following the paper Ref.15 we will find the set of the generalized Hermite-Gaussian beams on the base of generatrix functions as the particular solutions to the equations (31) and (32) in the form

$$\Psi_E^{(0)} = \frac{w_{xx} w_{xy}}{\sqrt{(w_{xx}^2 + i\zeta_x)(w_{xy}^2 + i\zeta_x)}} \exp\left(-\frac{x^2}{w_{xx}^2 + i\zeta_x} - \frac{y^2}{w_{xy}^2 + i\zeta_x}\right), \quad (45)$$

$$\Psi_H^{(0)} = \frac{w_{yx} w_{yy}}{\sqrt{(w_{yx}^2 + i\zeta_y)(w_{yy}^2 + i\zeta_y)}} \exp\left(-\frac{x^2}{w_{xy}^2 + i\zeta_y} - \frac{\frac{n_x^2}{n_y^2} y^2}{w_{yy}^2 + i\zeta_y}\right), \quad (46)$$

where $w_{xx}, w_{xy}, w_{yx}, w_{yy}$ are waists of the *E*- and *H*-beam at the $z=0$ plane (see Fig.1), $\zeta_{x,y} = \frac{2}{k_{x,y}} z$. Let us choose the differential operators in the form

$$\hat{H}_{x,E}^{(m)}(\beta_{x,E}) = (\beta_{x,E} - i\zeta_x)^m \frac{\partial^m}{\partial x^m}, \quad \hat{H}_{y,E}^{(m)}(\beta_{y,E}) = (\beta_{y,E} - i\zeta_x)^m \frac{\partial^m}{\partial y^n}, \quad (47)$$

$$\hat{H}_{y,H}^{(n)}(\beta_{y,H}) = (\beta_{y,H} - i\zeta_y)^n \frac{\partial^n}{\partial y^n}, \quad \hat{H}_y^{(n)}(\beta_{y,H}) = \left(\frac{n_y}{n_x}\right)^n (\beta_{y,H} - i\zeta_y)^n \frac{\partial^n}{\partial \overline{y}^n}, \quad (48)$$

Similar to that we can construct the operators $\hat{H}_{y,E}^{(n)}(\beta_{y,E})$ and $\hat{H}_{x,H}^{(m)}(\beta_{x,H})$, where $\beta_{x,E,H}, \beta_{y,E,H}$ stand for complex parameters., $\overline{y} = \frac{n_x}{n_y} y$. In particular, the operators have the important property:

$$\lim_{\beta_x \to \infty}\left(\frac{1}{\beta_x}\hat{H}_x^{(m)}\right) = \frac{\partial^m}{\partial x^m}, \quad \lim_{\beta_y \to \infty}\left(\frac{1}{\beta_y}\hat{H}_y^{(n)}\right) = \frac{\partial^n}{\partial y^n}. \quad (49)$$

The operators (47) and (48) commute with the operators $\hat{L}_E = \partial_x^2 + \partial_y^2 + i2k_x \partial_z$ and $\hat{L}_H = \partial_x^2 + \partial_y^2 + i2k_y \partial_z$ of the equations (31) and (32), respectively [15], and, consequently, the functions $\Psi_E^{(m,n)} = \hat{H}_E^{(m,n)} \Psi_E^{(0)}$ and $\Psi_H^{(m,n)} = \hat{H}_H^{(m,n)} \Psi_H^{(0)}$, where $\hat{H}_E^{(m,n)} = \hat{H}_{x,E}^{(m)} \hat{H}_{y,E}^{(n)}$, $\hat{H}_H^{(m,n)} = \hat{H}_{x,H}^{(m)} \hat{H}_{y,H}^{(n)}$ are also the solutions to the paraxial equations (31) and (32). Thus, the solutions to these paraxial equations in the form of the generalized Hermite-Gaussian beams are

$$\Psi_E^{(m,n)} = i^{m+n} \left( \frac{\beta_{x,E} - i\zeta_x}{w_{xx}^2 + i\zeta_x} \right)^{\frac{m}{2}} \left( \frac{\beta_{y,E} - i\zeta_x}{w_{xy}^2 + i\zeta_x} \right)^{\frac{n}{2}} \times$$
$$H_m \left( x \sqrt{\frac{\beta_{x,E} + w_{xx}^2}{(\beta_{x,E} - i\zeta_x)(w_{xx}^2 + i\zeta_x)}} \right) H_n \left( y \sqrt{\frac{\beta_{y,E} + w_{xy}^2}{(\beta_{y,E} - i\zeta_x)(w_{xy}^2 + i\zeta_x)}} \right) \Psi_E^{(0)} \quad (50)$$

$$\Psi_H^{(m,n)} = i^{m+n} \left( \frac{\beta_{x,H} - i\zeta_y}{w_{yx}^2 - i\zeta_y} \right)^{\frac{m}{2}} \left( \frac{\beta_{y,H} - i\zeta_y}{w_{yy}^2 - i\zeta_y} \right)^{\frac{n}{2}} \times$$
$$H_m \left( x \sqrt{\frac{\beta_{x,H} + w_{yx}^2}{(\beta_{x,H} - i\zeta_y)(w_{yx}^2 + i\zeta_y)}} \right) H_n \left( \frac{n_x}{n_y} y \sqrt{\frac{\beta_{y,H} + w_{yy}^2}{(\beta_{y,H} - i\zeta_y)(w_{yy}^2 + i\zeta_y)}} \right) \Psi_H^{(0)} \quad (51)$$

where we made use of the relationship:

$$\frac{\partial^m}{\partial x^m} e^{-x^2} = (-1)^m H_m(x) e^{-x^2}. \quad (52)$$

In case the parameters $\sqrt{\beta_{x,E}} = w_{xx}, \sqrt{\beta_{yE}} = w_{xy}, \sqrt{\beta_{x,H}} = w_{yx}, \sqrt{\beta_{y,H}} = w_{yy}$ the generatrix functions (50) and (51) are transformed into *standard Hermite-Gaussian beams*:

$$\Psi_E^{(m,n)} = i^{m+n} \left( \frac{w_{xx}^2 - i\zeta_x}{w_{xx}^2 + i\zeta_x} \right)^{\frac{m}{2}} \left( \frac{w_{xy}^2 - i\zeta_x}{w_{xy}^2 + i\zeta_x} \right)^{\frac{n}{2}} H_m \left( \sqrt{2} \frac{x}{|w_{xx}^2 + i\zeta_x|} \right) H_n \left( \sqrt{2} \frac{y}{|w_{xy}^2 + i\zeta_x|} \right) \Psi_E^{(0)}$$
$$, \quad (53)$$

$$\Psi_H^{(m,n)} = i^{m+n} \left( \frac{w_{yx}^2 - i\zeta_y}{w_{yx}^2 + i\zeta_y} \right)^{\frac{m}{2}} \left( \frac{w_{yy}^2 - i\zeta_y}{w_{yy}^2 + i\zeta_y} \right)^{\frac{n}{2}} H_m \left( \sqrt{2} \frac{x}{|w_{yx}^2 + i\zeta_y|} \right) H_n \left( \sqrt{2} \frac{\frac{n_x}{n_y} y}{|w_{yy}^2 + i\zeta_y|} \right) \Psi_H^{(0)}$$
$$. \quad (54)$$

In case the parameters $\beta_{x,E} \to \infty, \beta_{yE} \to \infty, \beta_{x,H} \to \infty, \beta_{y,H} \to \infty$, eqs (49) transforms the generatrix functions (50) and (51) into *the elegant Hermite-Gaussian beams*

$$\Psi_E^{(m,n)} = i^{m+n} \left(\frac{1}{\sqrt{w_{xx}^2 + i\zeta_x}}\right)^m \left(\frac{1}{\sqrt{w_{xy}^2 + i\zeta_x}}\right)^n H_m\left(\frac{x}{\sqrt{w_{xx}^2 + i\zeta_x}}\right) H_n\left(\frac{y}{\sqrt{w_{xy}^2 + i\zeta_x}}\right) \Psi_E^{(0)}$$
(55)

$$\Psi_H^{(m,n)} = i^{m+n} \left(\frac{1}{\sqrt{w_{yx}^2 + i\zeta_y}}\right)^m \left(\frac{1}{\sqrt{w_{yy}^2 + i\zeta_y}}\right)^n H_m\left(\frac{x}{\sqrt{w_{yx}^2 + i\zeta_y}}\right) H_n\left(\frac{\frac{n_x}{n_y} y}{\sqrt{w_{yy}^2 + i\zeta_y}}\right) \Psi_E^{(0)}$$
(56)

*2. Generalized elliptic beams in isotropic media*

Consider at first some important properties of the generalized Hermite-Gaussian beams in a homogeneous isotropic medium the attention was not focused at in Ref. 15-18. Let us form the function

$$Q = \left(\frac{w_y}{w_x}\right) \frac{1 + \left(\frac{\beta_x}{w_x^2}\right)}{1 + \left(\frac{\beta_y}{w_y^2}\right)},$$
(57)

where $w_{xx} = w_x, w_{yy} = w_y$, $\beta_{x,y,E} = \beta_{x,y}, \zeta_{x,y} = \zeta$, $z=0$ in eq. (50), that combines together all three basic parameters $\left(\frac{w_y}{w_x}\right)$, $\left(\frac{\beta_x}{w_x^2}\right)$ and $\left(\frac{\beta_y}{w_y^2}\right)$ of the generalized Hermite-Gaussian beam. Evolution of the beam cross-section on the plane $Q = const, w_y/w_x = 2, z = 0$ is shown in Fig.2. When $\left(\frac{\beta_x}{w_x^2}\right) = 1, \left(\frac{\beta_y}{w_y^2}\right) = 1$ the beam turns into the standard elliptic HG beam. It is the case that corresponds to the HG beams whose field distribution preserves along the *z*-axis up to the scale. All other cases are accompanied by vanishing the intrinsic edge dislocations out the plane *z=0*. When increasing the parameter $\left(\frac{\beta_y}{w_y^2}\right)$ while $\left(\frac{\beta_x}{w_x^2}\right) = 1$ (the beam perturbation follows the line $\left(\frac{\beta_x}{w_x^2}\right) = 1$), the edge dislocations along the *y*-axis vanish while ones along the *x*-axis remain without change for $z \neq 0$. Vice versa, when increasing the parameter $\left(\frac{\beta_x}{w_x^2}\right)$ while $\left(\frac{\beta_y}{w_y^2}\right) = 1$, the edge dislocations along the *x*-axis vanish while the dislocation pattern does not change along the *y*-axis. When $\left(\dfrac{\beta_x}{w_x^2}\right) \to 0$, but $\left(\dfrac{\beta_x}{w_x^2}\right) \neq 0$, $\left(\dfrac{\beta_y}{w_y^2}\right) = const$ the amplitude of the intermediate intensity oscillations decreases along the *x*-axis at the *z=0* plane while $\left(\dfrac{\beta_y}{w_y^2}\right) \to 0$ but $\left(\dfrac{\beta_y}{w_y^2}\right) \neq 0$, $\left(\dfrac{\beta_x}{w_x^2}\right) = const$ the oscillations decay along the *y*-axis. When $\left(\dfrac{\beta_x}{w_x^2}\right) \to \infty$ and $\left(\dfrac{\beta_y}{w_y^2}\right) \to \infty$ the generalized HG beam is transformed into the elegant HG beam. The ellipticity of spots in the intensity distribution of the beam is exclusively specified by the ration $\left(\dfrac{w_y}{w_x}\right)$ at the *z=0* plane.

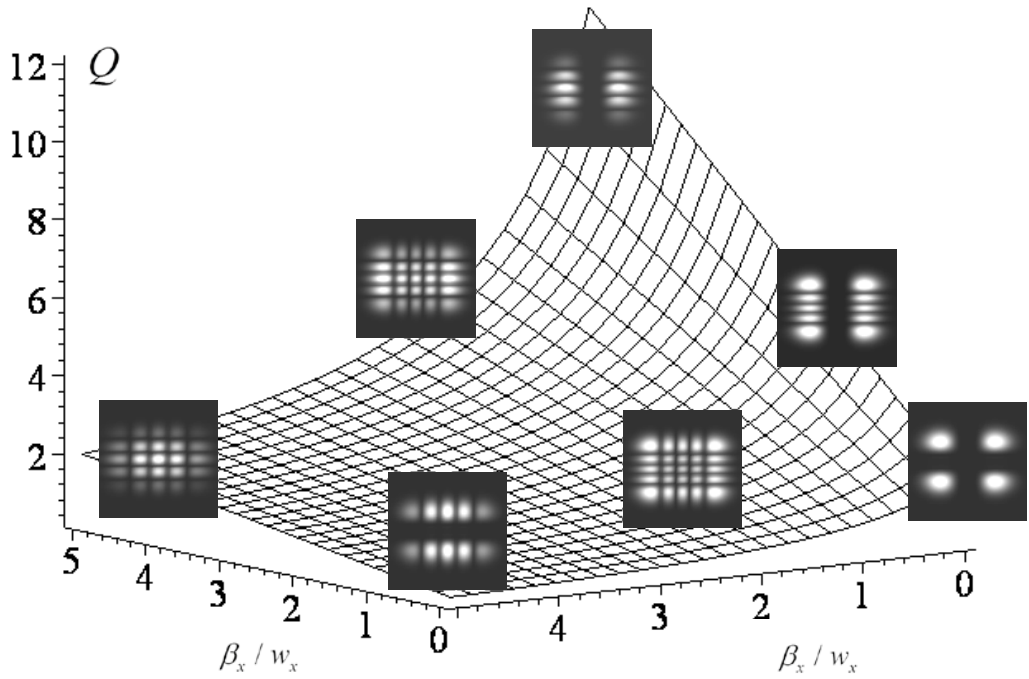

Fig.2 Transformations of the generalized Hermite-Gaussian beam at the plane
$Q = const$, $w_x / w_y = 0.5$, $z = 0$

    Let us form now the generalized Laguerre-Gaussian beam (LG beam) and analyze its properties. For this aim we will make use of the relation from Ref. 19 for the *z=0* plane:

$$\Psi^{(m,n)}(z=0) = (X+iY)^m L_n^m(X^2+Y^2) =$$
$$\frac{(-1)^{n+m}}{n!2^{2n+m}} \sum_{p=0}^{n}\sum_{q=0}^{m}(-i)^{m+q}\binom{n}{p}\binom{m}{q} H_{2p+q}(X) H_{2n+m-2p-q}(Y) \quad . \tag{58}$$

In order to find the field distribution at arbitrary $z$-plane, it is necessary the Fresnel integral operator $\hat{\mathcal{F}}$ to act on the expression (58) [18]:

$$\Psi^{(m,n)}(x,y,z) = \hat{\mathcal{F}}\left[\Psi^{(m,n)}(\xi,\eta,z=0)\right] =$$
$$\frac{k}{2\pi i z}\iint \exp\left(i\frac{k}{2z}|\mathbf{r}-\mathbf{\upsilon}|^2\right)\Psi^{(m,n)}(\xi,\eta)d\upsilon \quad , \tag{59}$$

where $\mathbf{\upsilon} = \mathbf{e_x}\xi + \mathbf{e_y}\eta$. Using eq.(50) for an isotropic medium we come to

$$\Psi^{(m,n)}(x,y,z) = \sum_{p=0}^{n}\sum_{q=0}^{m}(-i)^{m+q}\binom{n}{p}\binom{m}{q}\mathcal{H}_{2p+q}(X)\mathcal{H}_{2n+m-2p-q}(\Upsilon)\Psi^{(0)}(x,y,z), \tag{60}$$

where

$$\mathcal{H}_{2p+q}(X) = \left(\sqrt{\frac{\beta_x - i\zeta}{w_x^2 + i\zeta}}\right)^{2p+q} H_{2p+q}\left(x\sqrt{\frac{\beta_x + w_x^2}{(\beta_x - i\zeta)(w_x^2 + i\zeta)}}\right), \tag{61}$$

$$\mathcal{H}_{2n+m-2p-q}(\Upsilon) = \left(\sqrt{\frac{\beta_y - i\zeta}{w_y^2 - i\zeta}}\right)^{2n+m-2p-q} H_{2n+m-2p-q}\left(y\sqrt{\frac{\beta_y + w_y^2}{(\beta_y - i\zeta)(w_y^2 + i\zeta)}}\right). \tag{62}$$

where the normalized factor in eq. (60) is omitted.

For the high-order optical vortex embedded into the LG beam with $n=0$, the expression (58) is reduced to the form

$$(X+iY)^m = \sum_{q=0}^{m}(-i)^{m+q}\binom{m}{q}H_q(X)H_{m-q}(Y), \tag{63}$$

so that the wave function is now transformed into

$$\Psi^{(m,0)}(x,y,z) = \sum_{q=0}^{m}(-i)^{m+q}\binom{m}{q}\mathcal{H}_q(X)\mathcal{H}_{m-q}(\Upsilon)\Psi^{(0)}(x,y,z). \tag{64}$$

We have analyzed transformations of the standard elliptic LG beams with $\beta_x = w_x^2, \beta_y = w_y^2$. Typical evolution along the z-axis of the intensity and phase distributions in the elliptic LG beams with *n=4, m=4* and *n=0, m=4* is shown in Fig.3. Even a very small shift of the observation plane from the $z=0$ plane results in destroying the phase and intensity pattern as a whole in both LG beams. However, the LG$_{4,4}$ beam bearing both the four edge dislocations and the four-fold degenerate vortex forms more complex pattern than that shaped by the LG$_{0,4}$ beam with the only four-fold degenerate vortex. The elliptic perturbation out of the *z=0* plane just draws up the high-order vortex into the row from four singly charged vortices similar to that in Ince-Gaussian beams [11, 21], the vortex row being synchronously rotated when propagating the beam. More complex picture of the destroying phase singularities occurs in the elliptic LG$_{44}$ beam. The destroying process affects not only the degenerate centered

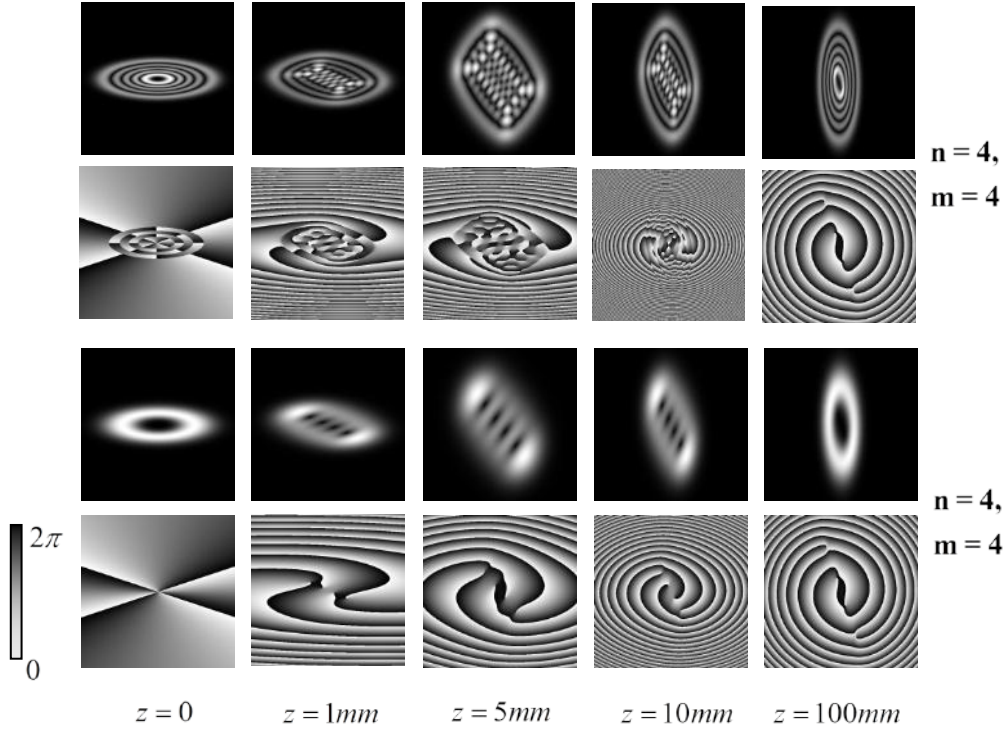

Fig. 3 Evolution of the intensity and phase distributions along the z-axis in the elliptic beam with $w_x = 30\,\mu m$, $w_y = 10\,\mu m$ and the refractive index $n = 2.3$, $\lambda = 0.6328\,\mu m$

vortex but also the ring dislocations. As the beam propagates, the ring dislocations begin going to pieces forming the complex cluster of optical vortices. At the far field, the singly charged vortices in the cluster gather together recovering the ring dislocations. However, the degenerate vortex state at the axis does not restore in contrast to those described in Ref. 18, 21. Our computer simulation have shown that the distance between the vortices decreases relative to a whole scale of the beam cross-section but the vortices are not centered at the axis, at least at the distance up to 10 m.

### 3. Circularly polarized elliptic beam in the crystal

The circularly polarized electric field of the elliptic beam is shaped as

$$E_+ = E_x - i E_y, \qquad E_- = E_x + i E_y. \tag{65}$$

The polarization components $E_x$ and $E_y$ are specified by eqs (33) and (40). In the normalized form they can be written as

$$E_x = \Psi_E e^{ik_x z}, \quad E_y = i\Psi_H e^{ik_y z} \tag{66}$$

so that eqs (65) have a form

$$E_+ = \Psi_E e^{ik_x z} + \Psi_H e^{ik_y z}, \qquad E_- = \Psi_E e^{ik_x z} - \Psi_H e^{ik_y z}, \tag{67}$$

We will deal here with only the case of the standard elliptic beams with the complex amplitudes specified by eqs (53) and (54). In order to the elliptic beam has only one circularly polarized component at the $z=0$ plane, for example $E_+$ component ($E_-(z=0)=0$) it is necessary to match the initial beam parameters, namely, $w_{xx} = w_{yx} = w_x$, $w_{xy} = w_y$ so that $\Psi_E(x,y,z=0) = \Psi_H(x,y,z=0)$:

$$\Psi_E^{(m,n)} = \left(\frac{\sigma_{xx}^*}{\sigma_{xx}}\right)^{\frac{m}{2}} \left(\frac{\sigma_{xy}^*}{\sigma_{xy}}\right)^{\frac{n}{2}} H_m\left(\sqrt{2}\frac{x}{w_x|\sigma_{xx}|}\right) H_n\left(\sqrt{2}\frac{y}{w_y|\sigma_{xy}|}\right) \Psi_E^{(0)}, \qquad (68)$$

$$\Psi_H^{(m,n)} = \left(\frac{\sigma_{yx}^*}{\sigma_{yx}}\right)^{\frac{m}{2}} \left(\frac{\sigma_{yy}^*}{\sigma_{yy}}\right)^{\frac{n}{2}} H_m\left(\sqrt{2}\frac{x}{w_x|\sigma_{yx}|}\right) H_n\left(\sqrt{2}\frac{y}{w_y|\sigma_{yy}|}\right) \Psi_H^{(0)}, \qquad (69)$$

where $\sigma_{xx} = 1 + iz/z_{xx}$, $\sigma_{xy} = 1 + iz/z_{xy}$, $\sigma_{yx} = 1 + iz/z_{yx}$, $\sigma_{yy} = 1 + iz/z_{yy}$, $z_{xx} = k_x w_x^2/2$, $z_{yx} = k_x w_y^2/2$, $z_{yx} = k_y w_x^2/2$, $z_{yy} = (k_x^2/k_y) w_y^2/2$ and

$$\Psi_E^{(0)} = \frac{1}{\sqrt{\sigma_{xx}\sigma_{xy}}} \exp\left(-\frac{x^2}{w_x\sigma_{xx}} - \frac{y^2}{w_y\sigma_{xy}}\right), \quad \Psi_E^{(0)} = \frac{1}{\sqrt{\sigma_{yx}\sigma_{yy}}} \exp\left(-\frac{x^2}{w_x\sigma_{yx}} - \frac{y^2}{w_y\sigma_{yy}}\right).$$
(70)

Also we will restrict themselves only to consideration of the high-order vortex beams with $n=0$ whose transverse field components go over the form

$$E_x^{(m,0)} = \Psi_E^{\{0\}} e^{ik_x z} \sum_{q=0}^m (-i)^{m+q} \binom{m}{q} \left(\frac{\sigma_{xx}^*}{\sigma_{xx}}\right)^{\frac{q}{2}} \left(\frac{\sigma_{xy}^*}{\sigma_{xy}}\right)^{\frac{m-q}{2}} H_q\left(\frac{\sqrt{2}\,x}{w_x|\sigma_{xx}|}\right) H_{m-q}\left(\frac{\sqrt{2}\,y}{w_y|\sigma_{xy}|}\right),$$
(71)

$$E_y^{(m,0)} = i\Psi_H^{\{0\}} e^{ik_y z} \sum_{q=0}^m (-i)^{m+q} \binom{m}{q} \left(\frac{\sigma_{yx}^*}{\sigma_{yx}}\right)^{\frac{q}{2}} \left(\frac{\sigma_{yy}^*}{\sigma_{yy}}\right)^{\frac{m-q}{2}} H_q\left(\frac{\sqrt{2}\,x}{w_x|\sigma_{yx}|}\right) H_{m-q}\left(\frac{\sqrt{2}\,y}{w_y|\sigma_{yy}|}\right).$$
(72)

The transformations of the intensity distribution along the crystal length z in such a vortex-beam are illustrated in Fig.4. Difference of the wave numbers and field distributions in the $E_x$ and $E_y$ components entails destructive interference that forms vortex dipoles breaking a straight row of the vortices originated from the centered degenerate singularity (see Fig.3). At the same time, the details of the transformation process depend on difference of the ordinary $n_x$ and extraordinary $n_y$ reflective indices of the crystal. A large difference $\Delta n = n_x - n_y$ results in different rotation angles of the vortex row in each linearly polarized components. Besides, the additional perturbation causes different deformations of the beam components along the *x*- and *y*-axes. The contributions of these processes to far field manifest themselves in the form of a partial separation of the beam ellipses shown in Fig.4a at the distance *z=1m*. The beam overlapping is negligibly small for a small value $\Delta n$ shown in Fig.4b. In far field, the basic vortices gather together near the beam axis as in Fig.3 for the ordinary beam

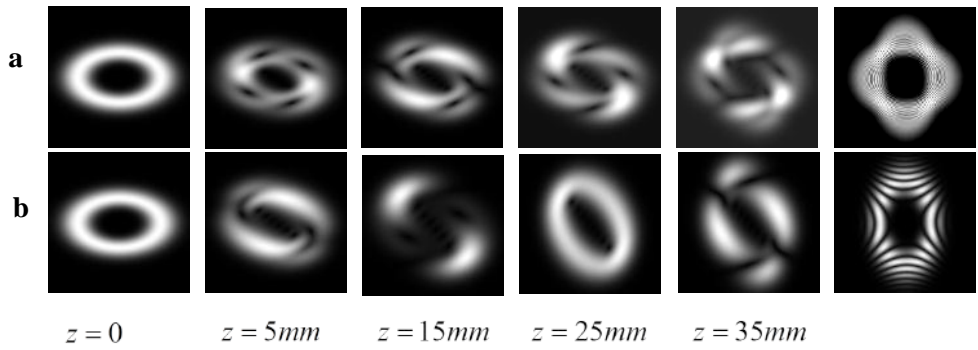

a

b

$z = 0 \quad z = 5mm \quad z = 15mm \quad z = 25mm \quad z = 35mm$

Fig. 4 Transformation of the intensity distributions in the $E_+$ component of the vortex beam with n=0 and m=10 in the crystal with (a) $n_x = 3, n_y = 2$, (b), $w_x = 50\,\mu m$, $w_y = 30\,\mu m$.

without forming the centered degenerate vortex while the additional vortices originated from the vortex dipoles annihilate with each other. Eventually the elliptically deformed conoscopic pattern of bright and dark hyperbolic lines is shaped both in the $E_+$ and $E_-$ components.

The partial bream overlapping is tightly coupled with homogeneity of the polarization states at the beam cross-section. The polarization inhomogeneity is associated with the vortex topological charge in the beam. The higher the vortex charge the stronger the polarization inhomogeneity manifests itself. The patterns shown in Fig.5 confirm this supposition. Indeed, centers of the splintered singly charged vortices in the linearly polarized $E_x$ and $E_y$ components do not coincide with each other forming a perturbed phase gradient at the vicinity of the vortex cores in the circularly polarized

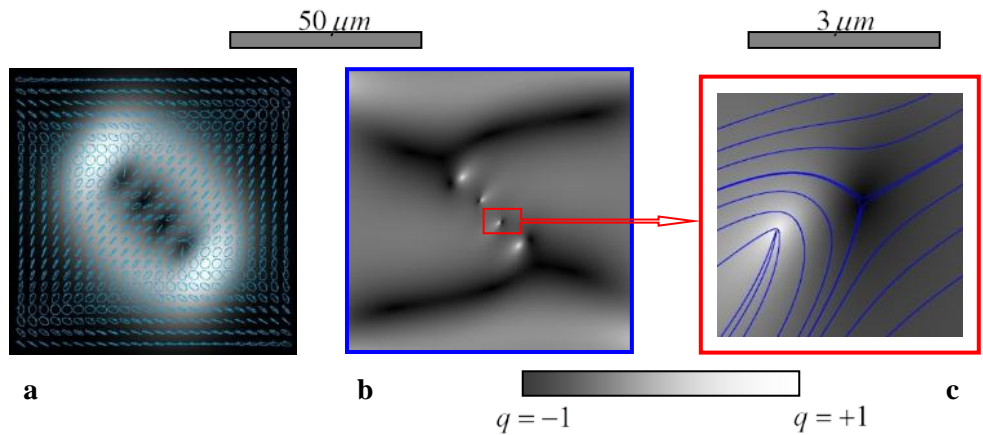

a  b  c

$q = -1 \qquad q = +1$

Fig.5 Polarization distribution in the vortex beam: (a) the distribution of the polarization states on the background of the intensity distribution of the vortex beam with $w_x = 10\,\mu m, w_y = 5\,\mu m$, $n = 0, m = 4$, $n_x = 2.5, n_y = 2$, $z = 1mm$, (b) the distribution of the polarization ellipticity $q$, (c) stream lines of the polarization states on the background of the ellipticity distribution

components. Fig. 5b illustrates the distribution of the polarization ellipticity $q = \pm b/a$ ($b$ and $a$ are the semiaxes of the polarization ellipse). Interchange of the bright and dark spots on the map of ellipticity in Fig.5b points out the sharp polarization transformations in these bean areas. The most visual characteristic of the polarization state distribution is the stream lines (the lines tangential to the large ellipse axis at each field point) [22]. The pattern of these lines shown in Fig.5c represents the vector dipole formed by the coupled pair of the star and lemon. When transmitting the elliptic beam the vector dipoles rotates around their mutual centers and roam about the beam cross-section changing radically the polarization state of the beam as a whole. At far field, the vector dipoles are drawn up together near the beam axis. However, they do not annihilate preserving some polarization disorder at any crystal length.

### IV. Spin angular momentum and conversion of the topological charge

Let us consider briefly aftermath of inhomogeneity of the polarization state over the beam cross-section and a beam perturbation along the y-axis on the example of the low-order vortex-beam. Our question is does the centered vortex change the sign of its topological charge when transmitting? To describe this process we rewrite the electric field of the $E_+$ and $E_-$ components in eqs (71) and (72) for *n=0, m=1* near the beam axis.

$$E_+^{(1,0)} \approx A_+ \frac{x}{w_x} + i\, a\, B_+ \frac{y}{w_y}, \quad E_-^{(1,0)} \approx A_- \frac{x}{w_x} + i\, a\, B_- \frac{y}{w_y}, \qquad (73)$$

where

$$A_\pm = \frac{e^{i\beta_x z}}{\sigma_{xx}\sqrt{\sigma_{xx}\sigma_{xy}}} \pm \frac{e^{i\beta_y z}}{\sigma_{yx}\sqrt{\sigma_{yx}\sigma_{yy}}}, \quad B_\pm = a\left(\frac{e^{i\beta_x z}}{\sigma_{xy}\sqrt{\sigma_{xx}\sigma_{xy}}} \pm \frac{e^{i\beta_y z}}{\sigma_{yy}\sqrt{\sigma_{yx}\sigma_{yy}}}\right), \qquad (74)$$

The parameter $a$ in eq. (73) adjusts the initial vortex core shape at the *z=0* plane. It is an independent parameter that can be inserted into the solutions (50) and (51) as the amplitude factor.

Description of the vortex state in each $E_+$ and $E_-$ components can be realized in terms of vectors fields $\psi_+ = \nabla_\perp E_+$ and $\psi_- = \nabla_\perp E_-$ [23, 24] that characterize the local phase distributions. The mathematical approach is based on the parameters similar to the Stokes parameters for polarization state of the beam:

$$S_0^{(\pm)} = |\nabla_\perp E_\pm|^2, \quad S_1^{(\pm)} = |\partial_x E_\pm|^2 - |\partial_y E_\pm|^2,$$
$$S_2^{(\pm)} = \partial_x E_\pm \partial_y E_\pm^* + \partial_x E_\pm^* \partial_y E_\pm, \quad S_3^{(\pm)} = i\left(\partial_x E_\pm^* \partial_y E_\pm - \partial_x E_\pm \partial_y E_\pm^*\right). \qquad (75)$$

However, the above parameters characterize the vortex shape rather than the polarization state. Deformation of the vortex core is described by the normalized $S_3^{(\pm)}$ in the form

$$\ell_z^{(\pm)} = \frac{i\left(\partial_x E_\pm^* \partial_y E_\pm - \partial_x E_\pm \partial_y E_\pm^*\right)}{|\nabla_\perp E_\pm|^2}. \qquad (76)$$

In the case of a simple vortex beam bearing the elliptically deformed singly charged vortex at the axis the value $\ell_z^{(\pm)}$ characterizes the orbital angular momentum of the beam.

In the more complex case the parameter $\ell_z^{(\pm)}$ describes the state of the vortex core: the modulus $\left|\ell_z^{(\pm)}\right|$ is of the ellipticity of the vortex core whereas its sign points out the sign of the vortex topological charge. The periodical curves plotted in Fig.6a refer to the conversion of the vortex ellipticity both in $E_+$ and $E_-$ components. The sharp peaks on the curve correspond to alternation of the positive and negative vortex topological charges. The beat length is defined by the simple relation: $\Lambda = \lambda / \left|n_x - n_y\right|$, where $\lambda$ is the wavelength in vacuum. In our case, it is about $\Lambda \approx 3.15\,\mu m$.

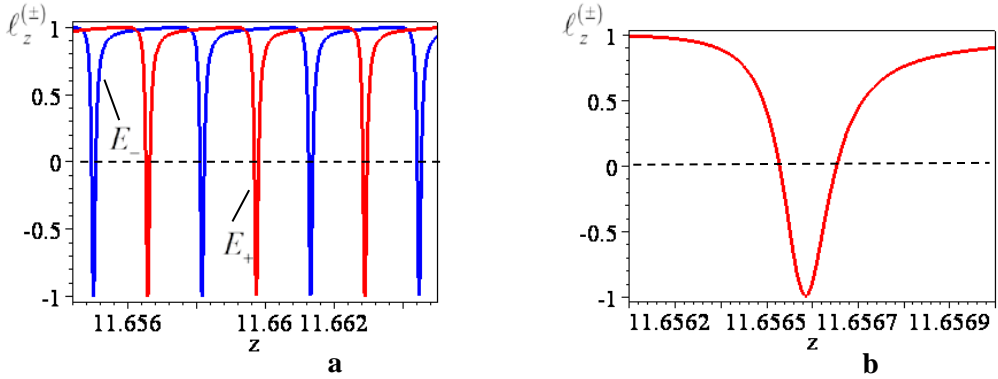

Fig.6 Periodical conversion of the vortex ellipticity along the crystal length z (cm) in the $E_+$ and $E_-$ components (a) ; the shape of the vortex-conversion line (b): $w_x = 2w_y = 10\,\mu m$, $n_x = 1.5, n_y = 1.7$, $a = 0.6$

The conversion peaks in the $E_+$ and $E_-$ components are shifted relative to each other at the distance $l = \Lambda / 2$. *But what is important: the vortex conversion occurs within very small ranges of the crystal length – a fraction of the wavelength.* Fig.6b illustrates the shape of the vortex conversion line. Its linewidth is about $\Delta l \approx 0.25\,\mu m$. The peaks of the vortex conversion match with the transformation of the polarization states from the right-hand to left-hand and vice-versa, i.e. *the sign of the vortex charge and the handedness of the polarization state change synchronically*. In the case of a plane wave such a polarization transformation is accompanied by the total energy transfer from one field components into another so that the vortex conversion cannot be observed in principle. However, the finite width of the singular beams results in the spatial field depolarization both in the vicinity of the vortex core and the beam as a whole. It is the depolarization process that can enable us to observe the vortex conversion.

The conversion process occurs non-instantly but is accompanied by a chainlet of the topological reactions. Typical pattern of the topological reactions is tracked with the help of evolution of the phase distribution shown in Fig.7. The regular phase spiral depicted in the first pattern of Fig.7 characterizes a nearly ideal state of the positively charged centered vortex. The vortex charge sign is described by the handedness of the red arrow in the figure. The perturbation of the phase spiral is caused by two topological dipoles born on the dotted line in the next figure. One of the newborn vortices comes nearer to the centered vortex causing the phase unfolding. Then the topological dipole is nucleated near the center again when transmitting the beam, the positively charged vortex remaining at the center while the negatively charged one being attracted to the other off-axis vortex to annihilate. The phase pattern in vicinity of the beam axis remains a nearly regular up to the next cycle of the vortex conversion. The length of the conversion cycle is a very short being about $\delta z \approx 0.32 \mu m$.

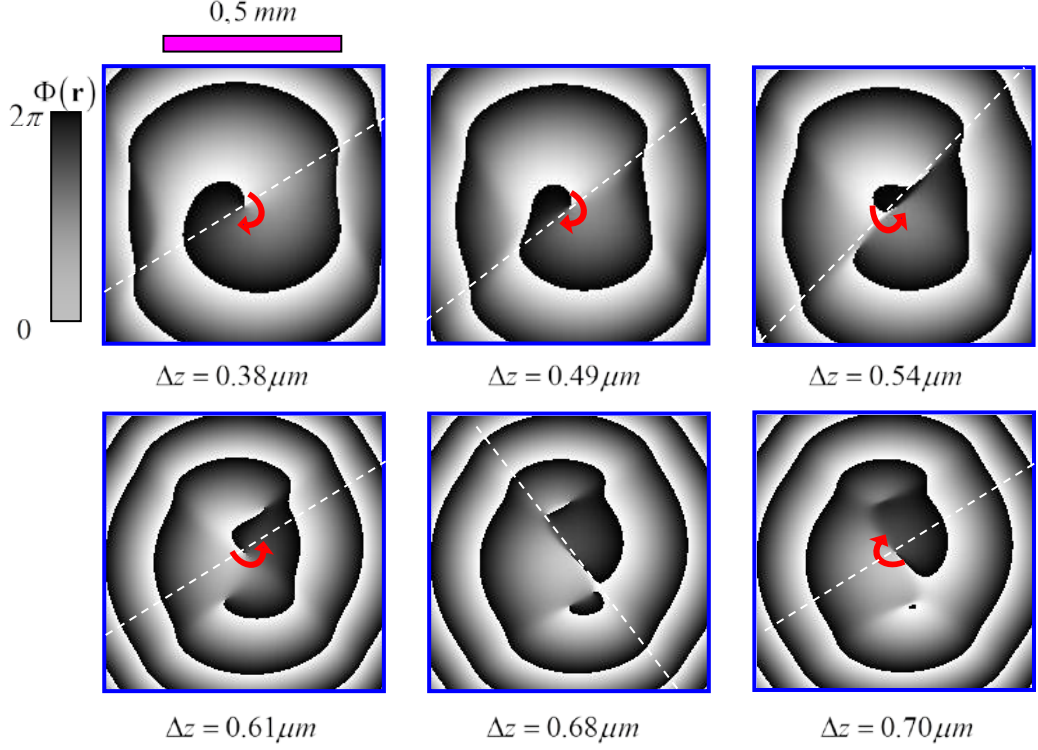

Fig.7 Evolution of the phase distribution $\Phi(\mathbf{r})$ of the $E_+$ component near the beam axis $r = 0$ along the crystal length $z = \bar{z} + \Delta z$ in vicinity of the vortex transformation for the beam and crystal parameters marked in Fig.6; $\bar{z} = 11.656 mm$

Perturbations of the polarization state at the beam cross-section for account of the elliptical deformation of the extraordinary beam imply the spatial light depolarization and the descent of the amplitude value of the spin angular momentum. Phase difference along the axial direction between the ordinary and extraordinary plane waves results in oscillating the polarization state from the right to left hand. However, the other plane waves in the angular spectra of the *E*- and *H*-beams acquire different values of the elliptic polarization so that the beam as a whole is depolarized. The spin angular momentum of the paraxial beam in a uniaxial crystal can be calculated as [25]

$$S_z = \frac{4 \operatorname{Im}\left(\int\limits_{-\infty}^{\infty} E_x E_y^* \, dx \, dy\right)}{\int\limits_{-\infty}^{\infty} \left(|E_x|^2 + |E_y|^2\right) dx \, dy} \qquad (77)$$

while the polarization degree is defined as

$$P = \frac{2\left|\int_{-\infty}^{\infty} E_x E_y^* \, dx\, dy\right|}{\int_{-\infty}^{\infty}\left(|E_x|^2 + |E_y|^2\right) dx\, dy}. \tag{78}$$

The results of our computer simulation of $S_z(z)$ and $P(z)$ are shown in Fig.8. The oscillations of the spin angular momentum (SAM) $S_z$ from $+1$ to $-1$ are typical only at the initial part of the crystal length. The positions of the SAM changing coincide with the positions of the vortex sign conversion and have the same beating length $\Lambda$. When increasing the crystal length the amplitude of oscillation comes gradually down to some utmost value unequal to zero. The envelope of the SAM oscillations is the polarization degree $P$. The utmost value of $P(z)$ depends on the beam waists $w_x$, $w_y$ and refractive indices of the crystal.

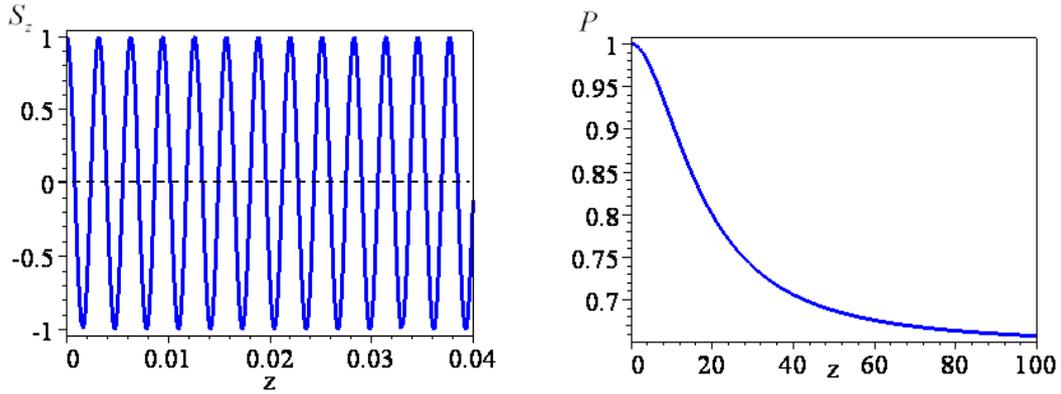

Fig.8 Evolution (a) of the spin angular momentum $S_z$ and (b) of the polarization state $P$ along the crystal length z (mm) in the vortex beam with parameters marked in Fig.6.

## V. Discussion and Conclusions

When analyzing the beam propagation perpendicular to the optical axis of the uniaxial crystal we have derived the corresponding equations and found their solutions in the form of the *generalized Hermite-Gaussian* beams. By manipulating these beams we have constructed the *generalized Laguerre-Gaussian* beams at the *z=0* plane and analyzed their evolution in a homogeneous isotropic medium. We have shown that the standard elliptic Laguerre-Gaussian (*seLG*) beam bearing the centered high order optical vortex at the initial plane *z=0* is destroyed when propagating in such a way that the centered high order vortex is split into a row of singly charged vortices while the ring dislocations are split into series of the topological dipoles. All vortices in this construction get mixed forming a complex combined singular beam. At far field, the ring dislocations recover a nearly regular structure while the singly charged vortices near the beam axis are not united into the centered high order optical vortex at any beam length.

We have traced the propagation of the circularly polarized *seLG* beam with the zero radial index $n=0$ bearing the *m*-charged centered vortex along the crystal, the ellipse axis of the beam cross-section being directed along the crystallographic axes. We have shown that the additional elliptic deformation of the extraordinary beam results in topological reactions that cannot favour recovering the initial singular structure. The degenerated centered vortex decays into a row of singly charged vortices when shifting slightly from the *z=0* plane as it takes place in free space. However, the additional geometrical perturbations in the extraordinary beam result in different rotation rates of the perturbed vortex rows in the ordinary and extraordinary beams. In the long run, the form of the resulting beam is radically distorted at far field. The beam loses its initial elliptic shape while the conoscopic pattern in the form of the distorted hyperbolic lines paints the beam components up to irrecognizability. It is such a complex picture of the field distribution in each circularly polarized component that causes the spatial depolarization of the vector singular beam.

We have predicted the conversion of the vortex topological charge in the singly charged elliptic vortex-beam similar to that in the astigmatic lens [12] or optical fibers [26]. However, there is a radical difference in these processes.

First of all, the vortex conversion in astigmatic lenses and optical fibers is associated either with different curvature of the lens surface along the major astigmatic axes or variations of the refractive index of the fiber cross-section. In our case we deal with homogenous medium. However, symmetry of the crystal birefringence causes the symmetry breakdown in the extraordinary beam spreading through the crystal changing the scale along one of the crystallographic axes. At the same time, the resulting elliptic deformation of the extraordinary beam provokes a perturbation of its helix wavefront. The destructive interference between the unperturbed ordinary beam and perturbed extraordinary beams entails a chainlet of topological reactions in vicinity of the beam axis and the periodical process of the vortex conversion at the beam axis.

In the second place, the singular beam with the converted vortex can propagate along indefinite large distance after the astigmatic lens while *the length-life of the converted beam in the crystal is considerably lesser or comparable with the wavelength*. This is conditioned by the wavefront shape of the perturbed extraordinary beam. By all appearances, it is the reason why this drastic phenomenon was not early revealed.

Different field distributions in the ordinary and extraordinary beams are also associated with peculiarities of the spin angular momentum of the singular beam. The oscillations of the spin angular momentum of the vortex-beam presented in the paper are accompanied by the spatial depolarization of the field that decreases the SAM amplitude. Notice that in case the beam propagates along the crystal optical axis any transformation of the SAM is compensated by the orbital angular momentum (OAM) in the form of the optical vortex generation so that the sum of the SAM and OAM is conserved. The axial symmetry of the crystal in this direction permits to divide the SAM and OAM from the mechanical angular momentum of the crystal medium so that any transformation of the SAM has an immediate impact on the OAM without implicating the crystal medium in the process. For example, a small beam inclination relative to the crystal optical axis entails the nonlocal lateral shift of the beam to compensate the access of the OAM [27] while the mechanical angular momentum of the medium is not involved.

The beam propagation perpendicular to the crystal optical axis breaks the former symmetry. Naturally, the vortex conversion in this case results in transforming the OAM. But now three processes take part in the phenomenon. The SAM and OAM are supplemented by the response of the crystal medium. It is the sum of these three processes that must reduce to conservation of the total angular momentum. Thus, the contribution of the vortex conversion to these three processes needs more detail theoretical and experimental investigations.

## Acknowledgement

A.Volyar thanks E. Abramochkin for the fruitful discussion on the mathematical background of the generalized Hermite-Laguerre-Gaussian beams.